\PassOptionsToPackage{unicode}{hyperref}
\PassOptionsToPackage{hyphens}{url}
\PassOptionsToPackage{dvipsnames,svgnames,x11names}{xcolor}
\documentclass[
12pt]{article}

\usepackage{amsmath,amssymb,amsthm} 
\usepackage{iftex}
\ifPDFTeX
\usepackage[T1]{fontenc}
\usepackage[utf8]{inputenc}
\usepackage{textcomp} 
\else 
\usepackage{unicode-math}
\defaultfontfeatures{Scale=MatchLowercase}
\defaultfontfeatures[\rmfamily]{Ligatures=TeX,Scale=1}
\fi
\usepackage{lmodern}
\ifPDFTeX\else  
\fi
\IfFileExists{upquote.sty}{\usepackage{upquote}}{}
\IfFileExists{microtype.sty}{
	\usepackage[]{microtype}
	\UseMicrotypeSet[protrusion]{basicmath} 
}{}
\makeatletter
\@ifundefined{KOMAClassName}{
	\IfFileExists{parskip.sty}{%
		\usepackage{parskip}
	}{
		\setlength{\parindent}{0pt}
		\setlength{\parskip}{6pt plus 2pt minus 1pt}}
}{
	\KOMAoptions{parskip=half}}
\makeatother
\usepackage{xcolor}
\makeatletter
\ifx\paragraph\undefined\else
\let\oldparagraph\paragraph
\renewcommand{\paragraph}{
	\@ifstar
	\xxxParagraphStar
	\xxxParagraphNoStar
}
\newcommand{\xxxParagraphStar}[1]{\oldparagraph*{#1}\mbox{}}
\newcommand{\xxxParagraphNoStar}[1]{\oldparagraph{#1}\mbox{}}
\fi
\ifx\subparagraph\undefined\else
\let\oldsubparagraph\subparagraph
\renewcommand{\subparagraph}{
	\@ifstar
	\xxxSubParagraphStar
	\xxxSubParagraphNoStar
}
\newcommand{\xxxSubParagraphStar}[1]{\oldsubparagraph*{#1}\mbox{}}
\newcommand{\xxxSubParagraphNoStar}[1]{\oldsubparagraph{#1}\mbox{}}
\fi
\makeatother

\usepackage{longtable,booktabs,array} 
\usepackage{calc} 
\usepackage{etoolbox}
\makeatletter
\patchcmd\longtable{\par}{\if@noskipsec\mbox{}\fi\par}{}{}
\makeatother
\IfFileExists{footnotehyper.sty}{\usepackage{footnotehyper}}{\usepackage{footnote}}
\makesavenoteenv{longtable}
\usepackage{graphicx}
\makeatletter
\def\maxwidth{\ifdim\Gin@nat@width>\linewidth\linewidth\else\Gin@nat@width\fi}
\def\maxheight{\ifdim\Gin@nat@height>\textheight\textheight\else\Gin@nat@height\fi}
\makeatother
\setkeys{Gin}{width=\maxwidth,height=\maxheight,keepaspectratio}
\makeatletter
\def\fps@figure{htbp}
\makeatother

\makeatletter
\@ifpackageloaded{caption}{}{\usepackage{caption}}
\AtBeginDocument{%
	\ifdefined\contentsname
	\renewcommand*\contentsname{Table of contents}
	\else
	\newcommand\contentsname{Table of contents}
	\fi
	\ifdefined\listfigurename
	\renewcommand*\listfigurename{List of Figures}
	\else
	\newcommand\listfigurename{List of Figures}
	\fi
	\ifdefined\listtablename
	\renewcommand*\listtablename{List of Tables}
	\else
	\newcommand\listtablename{List of Tables}
	\fi
	\ifdefined\figurename
	\renewcommand*\figurename{Figure}
	\else
	\newcommand\figurename{Figure}
	\fi
}
\@ifpackageloaded{float}{}{\usepackage{float}}
\floatstyle{ruled}
\@ifundefined{c@chapter}{\newfloat{codelisting}{h}{lop}}{\newfloat{codelisting}{h}{lop}[chapter]}
\floatname{codelisting}{Listing}

\makeatother
\makeatletter
\@ifpackageloaded{caption}{}{\usepackage{caption}}
\@ifpackageloaded{subcaption}{}{\usepackage{subcaption}}
\makeatother

\usepackage{dsfont} 
\usepackage{multirow}
\usepackage{siunitx}
\usepackage{enumitem} 

\ifLuaTeX
\usepackage{selnolig}  
\fi
\usepackage[]{natbib}
\usepackage{bookmark}

\IfFileExists{xurl.sty}{\usepackage{xurl}}{} 
\hypersetup{
	pdftitle={Fisher's Randomization Test for Causality with General Types of Treatments}, 
	pdfauthor={}, 
	pdfkeywords={Unconfoundedness, Sensitivity Analysis, Observational Studies, Potential Outcomes}, 
	colorlinks=true,
	linkcolor={blue},
	filecolor={Maroon},
	citecolor={Blue},
	urlcolor={Blue},
	pdfcreator={LaTeX via pandoc}}

\newtheorem{assumption}{Assumption}
\newtheorem{definition}{Definition}

\newtheorem{hypothesis}{Hypothesis}
\newtheorem{theorem}{Theorem}

\newtheorem{lemma}[]{Lemma}
\newtheorem{proposition}[]{Proposition}
\theoremstyle{definition}

\newcommand{\ci}{\perp\!\!\!\perp}

\newcommand{\anon}{1} 

\begin{document}
	
	\def\spacingset#1{\renewcommand{\baselinestretch}%
		{#1}\small\normalsize} 
	
	
	\if1\anon
	{
		\title{\bf Fisher's Randomization Test for Causality with General Types of Treatments} %
		\author{Zhen Zhong\thanks{
				The authors gratefully acknowledge Professor Donald B. Rubin and Professor Junni L. Zhang for discussion on core concepts, and Dr. Lu Deng and Ms. Jingjing Zhang for motivation with industrial applications in WeChat experimentation platform.}\hspace{.2cm}\\
			Faculty of Business and Economics, The University of Hong Kong}
		\maketitle
	} \fi
	
	\if0\anon
	{
		\bigskip
		\bigskip
		\bigskip
		\begin{center}
			{\LARGE\bf Fisher's Randomization Test for Causality with General Types of Treatments}
		\end{center}
		\medskip
	} \fi
	
	\bigskip
	\begin{abstract}
		We extend Fisher's randomization test (FRT) to test conditional independence between observed outcomes and treatments given covariates in both randomized experiments and observational studies, with no restriction on the variable type of treatments. Under a generalized unconfoundedness assumption, we provide causal identification for this hypothesis. Our approach requires neither the no-interference nor the positive overlap assumption, making it a widely applicable tool for detecting causal effects. A unique advantage of FRT lies in the separated roles of assignment and outcome models. The former, whether known from randomized experiments or estimated in observational studies, guarantees valid Type I error control at least asymptotically. The latter, even if misspecified, is used to construct optimal test statistics derived from Bayes factors. The synthesis of two classes of models through FRT yields a calibrated Bayesian procedure with desired frequentist properties. Recognizing that the generalized unconfoundedness assumption is untestable in observational studies, we develop a novel sensitivity analysis to assess the robustness of causal conclusions to unobserved confounding. Through a re-analysis of a panel dataset, we show how our methods can be integrated into a pipeline for observational causal inference.
	\end{abstract}
	
	\noindent%
	\vfill
	
	\newpage
	\spacingset{1.8}
	
	\section{Introduction}
	\label{sec:intro}
	Fisher's randomization test \citep{fisher1935design} is a unique tool for statistical inference in randomized experiments, valid under the sharp null hypothesis of no individual treatment effects. The validity does not rely on distributional assumptions about experimental outcomes, but only on the physical act of randomization by which the experiment is conducted. The sharp null hypothesis, however, is restrictive and has been dismissed as ``uninteresting and academic'' \citep{neyman1935statistical}. Recognizing this limitation, recent literature has focused on adapting FRT, often through carefully designed conditional events and test statistics, to address more meaningful hypotheses within finite-population settings \cite[e.g.][]{aronow2012general,athey2018exact,wu2021randomization,caughey2023randomisation}.
	
	In this paper, we adopt a perspective that combines sampling units from a larger population with the assignment of general types of treatments conditional on the sampled finite population \citep{rubin1978bayesian,abadie2020sampling}. We introduce a hypothesis of conditional independence between observed outcomes and treatments given covariates, and show that FRT remains valid under such a hypothesis. We build a unified causal identification framework for both randomized experiments and observational studies under the generalized unconfoundedness assumption. Notably, our framework requires neither the no-interference component of SUTVA \citep{rubin1980a} nor the positive overlap assumption \citep{rosenbaum1983assessing}. The circumvention of the latter assumption is particularly useful in observational causal inference. For categorical treatments, matching and trimming are standard techniques to enforce sufficient overlap \citep{imbens2015causal}, which would face the curse of dimensionality as the number of treatment levels grows. For continuous treatments, the positive overlap assumption is strong \citep{kennedy2017non}, and conceptually hard to check directly via finite samples \citep{wu2024matching}.
	
	A critical issue in FRT is the choice of test statistics. As the same data may contradict a hypothesis in different ways, Fisher deferred this choice to experimenter's domain knowledge \citep{lehmann1993fisher}. In this paper, we address this issue by considering alternative hypotheses expressed through a family of parametric outcome models. Echoing the wisdom that ``all models are wrong, but some are useful'' \citep{box1976science}, we do not assume these models are correctly specified. Instead, we use them to construct the test statistic derived from Bayes factors \citep{jeffreys1935some} for FRT. We theoretically prove that this choice maximizes Bayesian power among all test statistics at a fixed significance level. We also interpret such a choice for FRT as a frequentist calibration of Bayesian evidence \citep{rubin1984bayesian,little2006calibrated}.
	
	Unconfoundedness is an untestable assumption in observational studies \citep{imbens2015causal}. Since \cite{cornfield1959smoking}, sensitivity analysis has become a useful tool to examine how conclusions may change when unconfoundedness is violated at different levels. \cite{rosenbaum1995design} introduces a sensitivity analysis for FRT by bounding the difference between two estimated propensity odds ratios for a binary treatment. In this paper, we introduce a principled sensitivity analysis for FRT with no restrictions to the variable type of treatments. Since any unobserved variable that confounds the treatment-outcome relationship should be correlated with both, we propose to use the outcome as a proxy for unobserved confounders in the assignment model, and develop a computationally efficient method for its implementation. Through a re-analysis of the lottery data from \citet{imbens2001estimating}, we show how our methods can be integrated into a pipeline for observational causal inference. Besides standard results about statistical significance, our sensitivity analysis uncovers dynamic robustness of the causal effect to unobserved confounding over time.
	
	\noindent\textbf{Related Literature.}  Model-based FRTs have been used to test Fisher's sharp null hypothesis with noncompliance \citep{rubin1998more} and to test Neyman's weak null hypothesis in observational studies \citep{ding2023posterior}. A frequentist theory is lacking in these approaches. Another related body of literature is conditional permutation tests \citep[CPTs,][]{barber2015controlling,candes2018panning,berrett2020conditional,lazecka2023analysis}.
	\cite{zhang2023randomization} distinguish FRT from CPT by their presumptions as well as the hypotheses being tested, even if they are algorithmically similar. We find that FRT can test an identical hypothesis to CPT. However, CPT does not have causal identification and requires additional assumptions to ensure its validity.
	
	\section{Definitions, Assumptions and Hypotheses}
	\label{sec:pre}
	Consider an experiment with $n$ units indexed by $i=1,2,\ldots,n$. Let $\mathcal{W}$ denote the treatment space and $w\in\mathcal{W}$ a specific version of the treatment. In treatment-versus-control experiments, we can set $\mathcal{W}=\{0,1\}$ and let $w$ take the value of $0$ or $1$. In factorial experiments with $K$ factors, we can set $\mathcal{W}=\{0,1\}^{K-1}$ and let $w$ be a $(K-1)$-dimensional dummy variable. For continuous doses, we can set $\mathcal{W}$ to be an interval in the real line $\mathcal{R}$ and let $w$ represent a specific dose level.
	
	Let $\boldsymbol{w}=(w_1^{\top},\ldots,w_n^{\top})^{\top}$ denote a fixed assignment matrix, where each row $w_i\in\mathcal{W}$ specifies the treatment for unit $i$. The space $\mathcal{W}^n$ consists of all possible assignment matrices. Let $\boldsymbol{W}$ denote the realized assignment matrix, which is a random variable taking values in $\mathcal{W}^n$. The distribution of $\boldsymbol{W}$ is determined by an assignment mechanism. For example, in Bernoulli design, each entry $W_i$ is an independent Bernoulli variable.
	
	Consider units with pre-treatment covariates stored in an $(n\times d)$-dimensional covariate matrix $\boldsymbol{X}$, where the $i$-th row $X_i$ contains the covariates of unit $i$. Experimenters also balance covariates between treatment groups using stratification and/or rerandomization \citep{morgan2012rerandomization}. Consequently, the assignment mechanism can depend on these covariates. We denote the general probability density function of treatment assignment, conditional on covariates, as $f(\boldsymbol{W}\mid\boldsymbol{X})$.
	
	To define causal effects, we introduce the concept of potential outcomes \citep{neyman1923,rubin1974}. In its most general form, the potential outcome for unit \( i \) can be defined as follows.
	\begin{definition}
		\label{def:potential}
		For each unit \( i \), there exists a function
		\( Y_i(\cdot): \mathcal{W}^n \rightarrow \mathcal{R} \) that maps an assignment matrix \( \boldsymbol{w} \) to its potential outcome \( Y_i(\boldsymbol{w}) \); the observed outcome \( Y_i \) is equal to its potential outcome under the realized assignment, \(Y_i = Y_i(\boldsymbol{W})\).
	\end{definition}
	Sometimes, we can adopt the no interference part of the stable unit treatment value assumption (SUTVA, \citealp{rubin1980a}) to simplify analysis.
	\begin{assumption}[\textsc{No Interference}]
		\label{no_interference}
		The potential outcome for each unit $i$ does not depend on the treatment assigned to others, i.e., $Y_i(\boldsymbol{w}) = Y_i(w_i)$ for all $\boldsymbol{w} \in \mathcal{W}^n$.
	\end{assumption}
	Our general theory does not require Assumption \ref{no_interference}, but we will state explicitly when it is invoked.
	
	\subsection{The Science Table}
	The science table \citep{rubin2005causal} provides a conceptual framework for causal inference, which can be extended to Table \ref{tab:science} for general types of treatments. This table includes columns for unit-specific covariates $X_i$ and potential outcome functions $Y_i(\cdot)$. The ``Unit-Level Causal Contrasts'' column represents the comparison between a specific unit's potential outcomes under any two distinct assignment matrices, $\boldsymbol{w},\boldsymbol{w}' \in \mathcal{W}$, i.e., $Y_i(\boldsymbol{w}) \text{ vs. } Y_i(\boldsymbol{w}')$. The fundamental problem of causal inference \citep{holland1986statistics} exists in this general setting: for any given unit $i$, we can never observe both $Y_i(\boldsymbol{w})$ and $Y_i(\boldsymbol{w}^{\prime})$ simultaneously if $\boldsymbol{w} \neq \boldsymbol{w}'$. The last ``Summary Causal Effects'' column defines causal effects at the level of a collection of units $C$, which can represent the whole sample, or specific heterogeneous groups. A valid comparison must be made on a common set, i.e., $\{Y_i(\boldsymbol{w}),i\in C\}$ and $\{Y_i(\boldsymbol{w}^{\prime}),i\in C\}$.
	
	\begin{table}[htbp]
		\centering 
		\caption{``Science''--The Causal Estimand.}
		\label{tab:science} 
		\begin{tabular}{lcccc}
			\toprule
			\toprule
			&  & Potential & Unit-Level & Summary \\
			Units & Covariates & Outcome Functions & Causal Contrasts & Causal Effects \\
			\midrule
			1 &  $X_1$ & $Y_1(\cdot): \mathcal{W}^n \to \mathcal{R}$ & $Y_1(\boldsymbol{w}) \text{ \textit{vs.} } Y_1(\boldsymbol{w}')$ & \multirow{4}{*}{\parbox{3.5cm}{\centering Comparison of $Y_i(\boldsymbol{w})$ \textit{vs.} $Y_i(\boldsymbol{w}')$ across units $i$ for distinct $\boldsymbol{w}, \boldsymbol{w}'$.}} \\ 
			$\vdots$ & $\vdots$ & $\vdots$ & $\vdots$ & \\
			$i$ & $X_i$ & $Y_i(\cdot): \mathcal{W}^n \to \mathcal{R}$ & $Y_i(\boldsymbol{w}) \text{ \textit{vs.} } Y_i(\boldsymbol{w}')$ & \\
			$\vdots$ & $\vdots$ & $\vdots$ & $\vdots$ & \\
			$n$ & $X_n$ & $Y_n(\cdot): \mathcal{W}^n \to \mathcal{R}$ & $Y_n(\boldsymbol{w}) \text{ \textit{vs.} } Y_n(\boldsymbol{w}')$ & \\
			\bottomrule
			\bottomrule
		\end{tabular}
	\end{table}
	
	\subsection{Sampling and Assignment Mechanisms}
	Following \cite{rubin1978bayesian}, we identify two sources of uncertainty in randomized experiments. First, sampling uncertainty arises because $n$ units participating in an experiment are typically considered as a random sample drawn from a larger population of interest, $\mathcal{P}$, which can be a finite pool of $N \geq n$ units or an infinite super-population. Because the sampling is random, the covariates and potential outcome vector $\boldsymbol{Y}(\boldsymbol{w}) = (Y_1(\boldsymbol{w}), \dots, Y_n(\boldsymbol{w}))^{\top}$ for any fixed $\boldsymbol{w} \in \mathcal{W}^n$ are random variables. Consequently, all quantities defined in Table \ref{tab:science} can be viewed as random due to repeated sampling.
	
	Second, in a given randomized experiment, the realized treatment assignment matrix $\boldsymbol{W}$ is generated according to a known assignment mechanism $f(\boldsymbol{W}\mid \boldsymbol{X})$. \cite{rubin1978bayesian} notes that, with the assignment mechanism controlled in a randomized experiment, we can assume that for all $\boldsymbol{w}\in\mathcal{W}^n$,
	\begin{equation*}
		f(\boldsymbol{W} \mid \boldsymbol{Y}(\boldsymbol{w}), \boldsymbol{X}) = f(\boldsymbol{W} \mid \boldsymbol{X}),
	\end{equation*}
	or in \cite{dawid1979conditional}'s notation,
	\begin{equation}
		\label{unconfoundedness}
		\boldsymbol{W}\ci\boldsymbol{Y}(\boldsymbol{w})\mid\boldsymbol{X}.
	\end{equation}
	In observational studies, however, we need to treat equation \eqref{unconfoundedness} as an auxiliary assumption.
	
	Let $\boldsymbol{Y} = (Y_1, \dots, Y_n)^{\top}$ denote the vector of observed outcomes, where the $i$-th element is the potential outcome under the realized assignment: $Y_i = Y_i(\boldsymbol{W})$. The overall randomness in $\boldsymbol{Y}$ comes from both repeated sampling and assignment.
	\subsection{Hypotheses}
	Extending \cite{rubin1980a}, we can first state the sharp null hypothesis for the classic FRT.
	\begin{hypothesis}[\textsc{No Individual Treatment Effects}]
		\label{sharp}
		For $i=1,\ldots,n$, $Y_i(\boldsymbol{w})=Y_i(\boldsymbol{w}^{\prime})$ for all $\boldsymbol{w},\boldsymbol{w}^{\prime}\in\mathcal{W}^n$.
	\end{hypothesis}
	
	However, Hypothesis \ref{sharp} is overly restrictive and sample-specific. We instead focus on a hypothesis at any population level, stating that observed outcomes are conditionally independent of treatments given covariates.
	\begin{hypothesis}
		\label{ci_hyp}
		$\boldsymbol{Y}\ci\boldsymbol{W}\mid\boldsymbol{X}$.
	\end{hypothesis}
	Hypothesis \ref{sharp} implies Hypothesis \ref{ci_hyp}, but not vice versa unless the population of interest is the sample at hand. To illustrate this, consider a scenario under SUTVA with binary treatments where, for each unit $i$, the potential outcomes $Y_i(1)$ and $Y_i(0)$ are independently and identically distributed (i.i.d.) continuous random variables. Then Hypothesis \ref{ci_hyp} would hold because the distribution of the observed outcome $Y_i$ is identical regardless of whether treatment $W_i=0$ or $W_i=1$ is assigned. However, Hypothesis \ref{sharp} would generally not hold, as with continuous potential outcomes, $\Pr(Y_i(1)=Y_i(0))=0$.
	
	Hypothesis \ref{ci_hyp} can also be stated purely about the joint distribution of $(\boldsymbol{Y},\boldsymbol{W},\boldsymbol{X})$ without the background of randomized experiments. Indeed, conditional permutation tests (CPTs) often target the same hypothesis. However, the validity of CPT relies on additional assumptions, such as observations being i.i.d. \citep{barber2015controlling, candes2018panning, berrett2020conditional} or outcomes being discrete \citep{lazecka2023analysis}. Based on the known assignment mechanism, these additional assumptions are unnecessary for testing Hypothesis \ref{ci_hyp}. On the other hand, the generalized unconfoundedness assumption is essential for causal identification of Hypothesis \ref{ci_hyp}. We elaborate on these findings in the following sections.
	
	\section{Causal Identification and Inference with FRT}
	In this section, we develop a frequentist theory for testing Hypothesis \ref{ci_hyp} with FRT in randomized experiments, which can be readily generalized to observational studies.
	\subsection{Generalized Unconfoundedness and the Locality Principle}
	\cite{rosenbaum1983central} formalize a causal framework for both randomized experiments and observational studies by introducing the strong ignorability. Assuming SUTVA, let $\boldsymbol{Y}(\boldsymbol{0})=(Y_1(0),\ldots,Y_n(0))$ and $\boldsymbol{Y}(\boldsymbol{1})=(Y_1(1),\ldots,Y_n(1))$ denote the potential outcome vectors of $n$ units under control and treatment, respectively. \cite{rosenbaum1983central} defines the strongly ignorable assignment mechanism as
	\begin{align*}
		&\textsc{(Unconfoundedness) }\boldsymbol{W}\ci\left(\boldsymbol{Y}(\boldsymbol{0}),\boldsymbol{Y}(\boldsymbol{1})\right)\mid\boldsymbol{X},\\
		&\textsc{(Positive Overlap) }0<\Pr(W_i=1\mid\boldsymbol{X})<1, i=1,\ldots,n.
	\end{align*}
	For continuous treatments, \cite{keisuke2004continuous} introduce weak unconfoundedness as $ W_i \ci Y_i(w) \mid X_i$ for any treatment level $w\in\mathcal{W}$. Both these definitions of unconfoundedness are special cases of \eqref{unconfoundedness}, to which we refer as generalized unconfoundedness. Theorem \ref{ci_explanation} below establishes the causal identification of Hypothesis \ref{ci_hyp} for both randomized experiments and observational studies under such an assumption.
	\begin{theorem}
		\label{ci_explanation}
		Suppose that, for any fixed $\boldsymbol{w}\in\mathcal{W}^{n}$, the joint distribution of $(\boldsymbol{Y}(\boldsymbol{w}),\boldsymbol{W},\boldsymbol{X})$ admits a density function.
		Under generalized unconfoundedness \eqref{unconfoundedness}, the following two statements are equivalent:
		\begin{enumerate}
			\item $\boldsymbol{Y}\ci\boldsymbol{W}\mid\boldsymbol{X}$.
			\item For any fixed $\boldsymbol{w},\boldsymbol{w}^{\prime}\in\mathcal{W}^{n}$, if $f(\boldsymbol{w}\mid\boldsymbol{X})\cdot f(\boldsymbol{w}^{\prime}\mid\boldsymbol{X})>0$, then
			\begin{equation*}
				\boldsymbol{Y}(\boldsymbol{w})\mid\boldsymbol{X}\stackrel{d}{=}\boldsymbol{Y}(\boldsymbol{w}^{\prime})\mid\boldsymbol{X},
			\end{equation*}
			where $\stackrel{d}{=}$ means that two distributions are identical.
		\end{enumerate}
	\end{theorem}
	Following \cite{holland1986statistics}'s dictum that there can be ``no causation without manipulation'', Theorem \ref{ci_explanation} reveals that Hypothesis \ref{ci_hyp} compares potential outcome distributions under all manipulable treatments satisfying $f(\boldsymbol{w}\mid\boldsymbol{X}) > 0$, which we conceptualize as the locality principle. A closely related concept is the local average treatment effect \citep{angrist1996identification}, which defines the average causal effect for units whose treatment indicators could be manipulated by an instrument.
	
	In traditional hypothesis testing, a null hypothesis is often formulated as $H_0:\theta\in\Theta_0$ for a parametric model $f(\boldsymbol{Y}\mid\boldsymbol{W},\boldsymbol{X};\theta)$, where the locality principle is ignored. Following \cite{rosenbaum1983central}, observational causal inference typically adopts the positive overlap assumption (i.e., $f(w_i\mid\boldsymbol{X}) > 0$ for each unit $i$ and any $w_i\in\mathcal{W}$), which requires matching or trimming observations when lacking sufficient overlap. Because Hypothesis~\ref{ci_hyp} adheres to the locality principle, the positive overlap assumption is no longer needed, and the only crucial assumption is generalized unconfoundedness. The absence of the latter assumption can lead to Simpson's paradox \citep{simpson1951interpretation,dawid1979conditional}, where $\boldsymbol{Y}\ci\boldsymbol{W}\mid\boldsymbol{X}$ holds but $\boldsymbol{Y}\ci\boldsymbol{W}$ does not hold.
	
	\subsection{Type I Error Control}
	Under the sharp null Hypothesis \ref{sharp}, $\boldsymbol{Y}$ is fixed regardless of the specific treatment assignment. A test statistic $T(\boldsymbol{Y},\boldsymbol{W},\boldsymbol{X})$ is a real-valued, measurable function of the observed data. Conditional on the observed $(\boldsymbol{Y}, \boldsymbol{X})$, a Fisher's $p$-value is calculated as:
	\begin{equation}
		\label{fep}
		p=\Pr\left(T(\boldsymbol{Y},\boldsymbol{W}^{\text{rep}},\boldsymbol{X})\geq T(\boldsymbol{Y},\boldsymbol{W},\boldsymbol{X})\mid\boldsymbol{Y},\boldsymbol{X}\right),
	\end{equation}
	where the probability is taken over repeated assignment $\boldsymbol{W}^{\text{rep}}\sim f(\boldsymbol{W}^{\text{rep}}\mid\boldsymbol{X})$ identical to the original assignment mechanism $f(\boldsymbol{W}\mid\boldsymbol{X})$.
	
	Under Hypothesis \ref{sharp}, the Fisher's $p$-value is dominated by a uniform distribution in the sense that, for any fixed $\boldsymbol{Y},\boldsymbol{X}$ and $\alpha\in(0,1)$,
	\begin{equation}
		\label{typeI}
		\Pr(p\leq\alpha\mid\boldsymbol{Y},\boldsymbol{X})\leq\alpha,
	\end{equation}
	where the probability is taken over repeated assignment $\boldsymbol{W}\sim f(\boldsymbol{W}\mid\boldsymbol{X})$. Equation \eqref{typeI} implies that the classic FRT has valid Type I error control under Hypothesis \ref{sharp}, where randomness comes solely from the known treatment assignment mechanism. The following lemma tells that the same $p$-value also ensures valid Type I error control under Hypothesis \ref{ci_hyp}, where randomness comes from both repeated sampling and assignment.

	\begin{lemma}
		\label{lem:FRT}
		Suppose the joint distribution of observed data admits a probability density function $f(\boldsymbol{Y}, \boldsymbol{W}, \boldsymbol{X})$. Under Hypothesis \ref{ci_hyp}, the Fisher $p$-value defined in \eqref{fep} satisfies:
		$$ \Pr(p \leq \alpha) \leq \alpha\quad\text{ for }\alpha \in [0, 1],$$
		where the probability is taken over the joint distribution $f(\boldsymbol{Y}, \boldsymbol{W}, \boldsymbol{X})$.
	\end{lemma}
	Following \cite{neyman1923}'s seminal work, a growing literature \citep[e.g.][]{lin2013agnostic,li2017general} has focused on design-based approaches. These approaches, similar to ours, use only the known assignment mechanism to derive reference distributions for test statistics. However, they typically focus on causal estimands defined at the sample level (e.g., sample average treatment effects), whereas Hypothesis \ref{ci_hyp} can target larger populations of interest. Furthermore, many design-based frameworks are tailored to simple experimental designs like complete randomization, while FRT accommodates any probabilistic experimental design. The flexibility of FRT makes it easy to generalize beyond randomized experiments to observational studies, as we will discuss in Section \ref{sec:obs}.
	
	\section{Choosing Test Statistics: a Bayesian Approach}
	\label{sec:bayes}
	The concept of Type II error, defined as the frequency of falsely accepting the null hypothesis when a specific alternative hypothesis is true, was introduced by \citet{neyman1933ix}. However, \citet{fisher1955statistical} criticized this concept, arguing that it depends not only on the frequency with which alternative hypotheses are in fact true, but also greatly on how closely they resemble the null hypothesis. Building on the Bayesian hypothesis testing framework, we provide a new concept of Bayesian power to address this debate. This approach allows for incorporating experimenters' subjective knowledge about plausible alternative hypotheses into the FRT framework.
	
	\subsection{FRT Using Bayes Factors as Test Statistics}
	\label{subsec:frt_bf}
	\label{subsec:bayes_factor}
	Bayesian hypothesis testing, originally introduced by \cite{jeffreys1935some}, provides a framework for quantifying the evidence supporting a scientific theory using observed data \citep{kass1995bayes}. Let $f(z\mid\theta)$ be a probability density function defined on a general sample space $\mathcal{Z}$ parameterized by $\theta \in \Theta$. Consider two competing hypotheses defined by a partition of the parameter space: $H_0: \theta \in \Theta_0$ versus $H_1: \theta \in \Theta_1$, where $\Theta_0 \subset \Theta$ and $\Theta_1 = \Theta \setminus \Theta_0$. We begin by assigning prior distributions $\pi(\theta \mid H_0)$ on $\Theta_0$ and $\pi(\theta \mid H_1)$ on $\Theta_1$, which are typically derived from a global prior $\pi(\theta)$ over $\Theta$ restricted to $\Theta_0$ and $\Theta_1$, respectively. Given these priors and observed data $Z\in\mathcal{Z}$, we then calculate the Bayes factor as
	\begin{equation*}
		\frac{f(Z\mid H_1)}{f(Z\mid H_0)}=\frac{\int_{\Theta_1} f(Z\mid\theta)\pi(\theta \mid H_1)d\theta}{\int_{\Theta_0} f(Z\mid\theta)\pi(\theta \mid H_0)d\theta}.
	\end{equation*}
	
	Let $\delta(z)$ denote a randomized decision rule, which is a measurable function defined on $\mathcal{Z}$ and takes values in $[0,1]$. Given the observed data $Z\in\mathcal{Z}$, we would reject $H_0$ if a Bernoulli trial with probability $\delta(Z)$ turns out to be $1$. The Bayesian Type I error and Bayesian power are defined, respectively, as
	\begin{align*}
		\alpha_B &= \Pr(\delta=1 \mid H_0) = \int \delta(z) f(z\mid H_0)dz, \\
		\beta_B &= \Pr(\delta=1 \mid H_1) = \int \delta(z) f(z\mid H_1)dz.
	\end{align*}
	When the frequentist Type I error rate is controlled at a fixed level $\alpha$, we have $\alpha_B$ equal to $\alpha$ because
	\begin{equation*}
		\alpha_B= \int \delta(z) f(z\mid \theta)\pi(\theta\mid H_0)dzd\theta=\int\alpha\pi(\theta\mid H_0)d\theta=\alpha.
	\end{equation*}
	On the other hand, Bayesian power connects to the posterior probability of $H_1$ conditional on rejecting $H_0$. Let $\Pr(H_0)>0$ and $\Pr(H_1)>0$ be prior probabilities assigned to $H_0$ and $H_1$, respectively, such that $\Pr(H_0) + \Pr(H_1) = 1$. According to the Bayes' rule,
	\begin{equation*}
		\Pr(H_1\mid\delta=1)=\frac{\Pr(\delta=1\mid H_1)\Pr(H_1)}{\Pr(\delta=1)}=\frac{\Pr(\delta=1\mid H_1)\Pr(H_1)}{\Pr(\delta=1\mid H_0)\Pr(H_0)+\Pr(\delta=1\mid H_1)\Pr(H_1)}.
	\end{equation*}
	Therefore, with fixed $\Pr(H_0)$, $\Pr(H_1)$, and $\alpha_B$, maximizing $\beta_B$ is equivalent to maximizing the posterior probability of $H_1$ conditional on rejecting the null.
	
	Focusing on the problem of testing Hypothesis \ref{ci_hyp}, we consider a family of outcome models specified as $\mathcal{L}=\{f(\boldsymbol{Y}\mid\boldsymbol{W},\boldsymbol{X};\theta)\mid\theta\in\Theta\}$. Let $\Theta_0\subset\Theta$ denote the subspace for which Hypothesis \ref{ci_hyp} holds. The Bayes factor can be calculated as
	\begin{equation*}
		T_{\text{BF}}(\boldsymbol{Y},\boldsymbol{W},\boldsymbol{X}) = \frac{f(\boldsymbol{Y},\boldsymbol{W},\boldsymbol{X}\mid H_1)}{f(\boldsymbol{Y},\boldsymbol{W},\boldsymbol{X}\mid H_0)} = \frac{\int_{\Theta_1} f(\boldsymbol{Y}|\boldsymbol{W},\boldsymbol{X}; \theta)\pi(\theta \mid H_1) d\theta}{\int_{\Theta_0} f(\boldsymbol{Y}|\boldsymbol{X}; \theta)\pi(\theta \mid H_0) d\theta}.
	\end{equation*}
	A standard approach in Bayesian hypothesis testing is to compare Bayes factors against pre-defined thresholds \citep{jeffreys1961theory, kass1995bayes}. However, this approach is subjective for practical decision-making. Seeking to reconcile Bayes/non-Bayes thinking \citep{good1992bayes, berger2003could, robnik2022statistical}, we propose to use $T_{\text{BF}}$ as the test statistic in a FRT procedure. As shown in Lemma \ref{lem:FRT}, FRT provides valid Type I error control under Hypothesis \ref{ci_hyp}, regardless of whether the outcome models are correctly specified.

	We now theoretically justify that FRT using $T_{\text{BF}}$ is optimal among all choices of test statistics. Because the distribution of Fisher's $p$-value can be non-uniform, a deterministic decision rule that rejects Hypothesis \ref{ci_hyp} when $p\leq\alpha$ may have a lower false rejection rate than $\alpha$. To achieve an exact significance level, we induce a randomized decision rule. For fixed $\boldsymbol{Y},\boldsymbol{X}$, denote the cumulative distribution function as $F(t)=\Pr(p\leq t\mid\boldsymbol{Y},\boldsymbol{X})$. Let $\alpha^{+} = \sup \left\{ t \in \mathcal{R} \mid F(t) \leq \alpha \right\}$. Since $F(\alpha)\leq\alpha\leq F(\alpha^{+})$, a randomized decision rule can be defined as
	\begin{equation}
		\label{rd}
		\delta(\boldsymbol{Y},\boldsymbol{W},\boldsymbol{X})=\begin{cases}
			1, \text{ if }p<\alpha^{+},\\
			q, \text{ if }p=\alpha^{+},\\
			0, \text{ if }p>\alpha^{+},
		\end{cases}
	\end{equation}
	where $q$ is determined such that $\Pr(\delta=1\mid\boldsymbol{Y},\boldsymbol{X})=\alpha$. Let $\Delta_{\alpha}$ denote the class of all such randomized decision rules based on FRT at the significance level $\alpha$.
	
	\begin{theorem}
		\label{theorem:bayes_factor}
		Given the prior distribution $\pi$ and a family $\mathcal{L}$ of likelihoods, the decision rule generated by FRT using $T_{\text{BF}}$ as the test statistic achieves the maximum Bayesian power within $\Delta_{\alpha}$.
	\end{theorem}
	
	\subsection{Frequentist Calibration for Bayesian Evidence}
	The calculation of Fisher's $p$-value depends solely on the ordering of the observed test statistic within its reference distribution. Consequently, any strictly monotonic transformation of a given test statistic will yield an identical $p$-value. In practice, a computationally tractable statistic can be constructed by picking any $\theta_0\in\Theta_0$ and calculating the negative logarithm of posterior density at $\theta_0$:
	\begin{equation*}
		T_{\text{pos}}(\boldsymbol{Y},\boldsymbol{W},\boldsymbol{X}) = -\log\left(f(\theta_0 \mid \boldsymbol{Y},\boldsymbol{W},\boldsymbol{X})\right).
	\end{equation*}
	The next proposition tells that it is equivalent to using $T_{\text{BF}}$ or $T_{\text{pos}}$ as a test statistic for calculating Fisher's $p$-value.
	\begin{proposition}
		\label{proposition:sufficiency}
		The Fisher's $p$-value remains identical using either $T_{\text{BF}}$ or $T_{\text{pos}}$ as a test statistic.
	\end{proposition}
	In practice, we can use Laplace's approximation to calculate $T_{\text{pos}}$ when the analytic form of $f(\theta\mid \boldsymbol{Y},\boldsymbol{W},\boldsymbol{X})$ is unavailable, which does not hurt Type I error control of FRT.
	
	FRT using $T_{\text{pos}}$ can also be viewed as a calibration of the fully Bayesian approach. A common ad hoc practice for Bayesian hypothesis testing involves constructing a $(1-\alpha)$ highest posterior density (HPD) region, $C_{1-\alpha} = \{\theta \in \Theta \mid f(\theta \mid \boldsymbol{Y},\boldsymbol{W},\boldsymbol{X}) > t \}$, where $t$ is chosen such that $\int_{C_{1-\alpha}} f(\theta \mid \boldsymbol{Y},\boldsymbol{W},\boldsymbol{X}) d\theta = 1-\alpha$. The null hypothesis $H_0: \theta = \theta_0$ would then be rejected if $\theta_0 \notin C_{1-\alpha}$.
	One can also calculate the posterior probability of the region where the density is less than or equal to that at $\theta_0$:
	\begin{equation*}
		p_{\text{Bayes}} = \Pr(f(\theta \mid \boldsymbol{Y},\boldsymbol{W},\boldsymbol{X}) \le f(\theta_0 \mid \boldsymbol{Y},\boldsymbol{W},\boldsymbol{X}) \mid \boldsymbol{Y},\boldsymbol{W},\boldsymbol{X}),
	\end{equation*}
	or in terms of $T_{\text{pos}}$:
	\begin{equation*}
		p_{\text{Bayes}} = \Pr\left(-\log(f(\theta \mid \boldsymbol{Y},\boldsymbol{W},\boldsymbol{X})) \ge T_{\text{pos}}(\boldsymbol{Y},\boldsymbol{W},\boldsymbol{X}) \mid \boldsymbol{Y},\boldsymbol{W},\boldsymbol{X}\right).
	\end{equation*}
	The decision rule to reject $H_0: \theta = \theta_0$ when $\theta_0 \notin C_{1-\alpha}$ is equivalent to $p_{\text{Bayes}}\leq\alpha$.
	
	A key difference between FRT using $T_{\text{pos}}$ and the fully Bayesian approach lies in the reference distribution. FRT calibrates $T_{\text{pos}}$ against the randomization distribution generated by the assignment mechanism, while the fully Bayesian approach calibrates it against the posterior distribution of $\theta$. As noted in the Bayesian inference literature \cite[e.g.,][]{rubin1984bayesian,little2006calibrated,li2023bayesian}, appropriate frequency calculations are essential to link Bayesian evidence to real-world practice. Since the assignment mechanism describes the physical act of randomization, it provides a more reasonable basis for calibrating Bayesian evidence.
	
	\section{Generalization to Observational Studies}
	\label{sec:obs}
	Since Lemma \ref{lem:FRT} relies only on the joint distribution of observed data, it can be directly generalized to observational studies. While the true assignment mechanisms are generally unknown in observational studies, we can model them using observed treatments and covariates. Consider a family of assignment models $\{f(\boldsymbol{W} \mid \boldsymbol{X}; \psi) : \psi \in \Psi\}$. Let $\widehat{p}$ be Fisher's $p$-value computed under $\boldsymbol{W}^{\text{rep}} \sim f(\boldsymbol{W}^{\text{rep}} \mid \boldsymbol{X}; \widehat{\psi})$, where $\widehat{\psi}$ is estimated, e.g., from maximum likelihood. The next Lemma tells that $\widehat{p}$ guarantees asymptotically valid Type I error control under Hypothesis \ref{ci_hyp}, provided the assignment mechanism is consistently estimated.
	\begin{lemma}
		\label{lem:FRT_asmp}
		Suppose that the joint distribution of observed data admits a density function $f(\boldsymbol{Y},\boldsymbol{W},\boldsymbol{X};\psi)=f(\boldsymbol{Y}\mid\boldsymbol{W},\boldsymbol{X})f(\boldsymbol{W}\mid\boldsymbol{X};\psi)f(\boldsymbol{X})$, which is continuous at $\psi_0\in\Psi$. Under Hypothesis \ref{ci_hyp}, if $\widehat{\psi} \xrightarrow{p} \psi_0$, then
		\begin{equation*}
			\limsup_{n\rightarrow\infty} \Pr(\widehat{p} \leq \alpha) \leq \alpha\quad\text{ for }\alpha \in (0, 1),
		\end{equation*}
		where the probability is taken over the joint distribution $f(\boldsymbol{Y}, \boldsymbol{W}, \boldsymbol{X};\psi_0)$.
	\end{lemma}
	For Bayesian inference, let $\widehat{p}_{\text{pos}}$ be Fisher's $p$-value computed under $\boldsymbol{W}^{\text{rep}} \sim \int f(\boldsymbol{W}^{\text{rep}} \mid \boldsymbol{X}; \psi)f(\psi \mid \boldsymbol{W}, \boldsymbol{X}) d\psi$. As noted by \cite{ding2023posterior}, $\widehat{p}_{\text{pos}}$ is also a posterior predictive $p$-value in the sense of \cite{meng1994posterior}. The frequentist Type-I error control for $\widehat{p}_{\text{pos}}$ can be established similarly to $\widehat{p}$ in combination with the Bernstein-von Mises theorem.
	
	FRT is a convenient approach for detecting causal effects in observational studies, because its null hypothesis follows the locality principle, making it unnecessary to enforce sufficient overlap in covariate distributions for each treatment level. However, two kinds of biases can arise when conducting FRT with an estimated assignment mechanism. The first one is the estimation bias, which occurs if the true assignment mechanism is not included in the model space $\{f(\boldsymbol{W}\mid\boldsymbol{X};\psi)\mid\psi\in\Psi\}$, or the assignment model is overfitted or poorly estimated. There are many ways to reduce this kind of bias, such as using flexible model specification, conducting cross validation, and carefully examining model fit diagnostics.
	
	Another kind of bias comes from possible violation of generalized unconfoundedness, leading to the existence of unobserved confounding. \cite{cinelli2020making} develop a suite of tools for sensitivity analysis using the omitted variable bias (OVB) framework. However, the causal interpretation of the OVB framework is tempting, as it requires the ``true'' outcome model to be a linear structural equation. Another line of research focuses on model-based sensitivity analysis for well-defined causal estimands \cite[e.g., ][]{rosenbaum1983assessing,imbens2003sensitivity,dorie2016flexible}. In the case of FRT, sensitivity analysis can be conducted solely for assignment models, as the (asymptotic) validity of FRT does not rely on outcome model specification.
	
	Let $U_i$ denote an unobserved covariate for unit $i$ and $\boldsymbol{U}$ denote these for all units. In the presence of $\boldsymbol{U}$, let us assume that generalized unconfoundedness holds conditional on $\boldsymbol{X}$ and $\boldsymbol{U}$, i.e.,
	\begin{equation}
		\label{unconfoundedness_unobs}
		\boldsymbol{W}\ci\boldsymbol{Y}(\boldsymbol{w})\mid\{\boldsymbol{X},\boldsymbol{U}\}.
	\end{equation}
	If $\boldsymbol{U}$ were observed, the following hypothesis would have causal identification according to Theorem \ref{ci_explanation}.
	\begin{hypothesis}
		\label{ci_unobs_hyp}
		$\boldsymbol{Y}\ci\boldsymbol{W}\mid\{\boldsymbol{X},\boldsymbol{U}\}.$
	\end{hypothesis}

	\cite{rosenbaum1995design} introduces a sensitivity analysis for FRT with binary treatments under SUTVA, where a logit form links $\pi_{i}=\Pr(Z_i=1\mid X_i,U_i)$ to $(X_i,U_i)$ and $\zeta^{W}$ as
	\begin{equation*}
		\log\left(\frac{\pi_{i}}{1-\pi_{i}}\right)=\kappa(X_i)+\zeta^{W} U_i, \quad\text{with }0\leq U_i\leq1,
	\end{equation*}
	for $\kappa(\cdot)$ an unknown function. To carry out the sensitivity analysis, all $U_i$'s are set to  $1$, and a sequence of Fisher's $p$-values is calculated by altering $\zeta^{W}$ to different levels. Insignificant Fisher's $p$-values at certain levels of $\zeta^W$ suggest that the treatment effect is explained away by unobserved confounding.
	
	Alternatively, we can leverage the observed outcomes as a proxy for the unobserved covariates to carry out a sensitivity analysis. This is preferred to directly modeling $\boldsymbol{U}$, which is, by definition, unobserved. If $\boldsymbol{U}$ is a confounder, it correlates with both $\boldsymbol{W}$ and $\boldsymbol{Y}$ conditional on $\boldsymbol{X}$. Thus, $\boldsymbol{Y}$ serves as a proxy for $\boldsymbol{U}$ in the assignment model.
	
	This approach can be formalized by considering two extended classes of assignment and outcome models under Hypothesis \ref{ci_unobs_hyp}, $\{f(\boldsymbol{W}\mid\boldsymbol{X},\boldsymbol{U};\psi,\zeta^{W}):\psi\in\Psi,\zeta^{W}\in\mathcal{R}\},\quad\{f(\boldsymbol{Y}\mid\boldsymbol{X},\boldsymbol{U};\theta,\zeta^{Y}):\theta\in\Theta,\zeta^{Y}\in\mathcal{R}\}$, where $\psi,\theta$ can be estimated given observed data, and $\zeta^{W},\zeta^{Y}$ are sensitivity parameters quantifying the relationship between $\boldsymbol{U}$ and $\boldsymbol{W},\boldsymbol{Y}$ conditional on $\boldsymbol{X}$, respectively.
	Using the conditional distribution $f(\boldsymbol{U}\mid\boldsymbol{Y},\boldsymbol{X};\theta,\zeta^{Y})\propto f(\boldsymbol{Y}\mid\boldsymbol{X},\boldsymbol{U};\theta,\zeta^{Y})f(\boldsymbol{U}\mid\boldsymbol{X})$, we integrate out $\boldsymbol{U}$ from the assignment model to obtain the conditional distribution for $\boldsymbol{W}$ as
	\begin{equation}
		\label{post_pred_W}
		f(\boldsymbol{W}\mid\boldsymbol{Y},\boldsymbol{X};\psi,\theta,\zeta^{W},\zeta^{Y})=\int f(\boldsymbol{W}\mid\boldsymbol{X},\boldsymbol{U};\psi,\zeta^{W}) f(\boldsymbol{U}\mid\boldsymbol{Y},\boldsymbol{X};\theta,\zeta^{Y})d\boldsymbol{U}.
	\end{equation}
	Using the marginal distributions \begin{align}
		\label{marginal_W}
		&f(\boldsymbol{W}\mid\boldsymbol{X};\psi,\zeta^{W})=\int f(\boldsymbol{W}\mid\boldsymbol{X},\boldsymbol{U};\psi,\zeta^{W})f(\boldsymbol{U}\mid\boldsymbol{X})d\boldsymbol{U},\\
		\label{marginal_Y}
		&f(\boldsymbol{Y}\mid\boldsymbol{X};\theta,\zeta^{Y})=\int f(\boldsymbol{Y}\mid\boldsymbol{X},\boldsymbol{U};\theta,\zeta^{Y})f(\boldsymbol{U}\mid\boldsymbol{X})d\boldsymbol{U},
	\end{align}
	we can estimate $\left(\widehat{\psi},\widehat{\theta}\right)$ for $\left(\psi,\theta\right)$. We then calculate Fisher's $p$-value $\widehat{p}(\zeta^{W},\zeta^{Y})$ for testing Hypothesis \ref{ci_unobs_hyp} using the reference distribution
	\begin{equation*}
		\boldsymbol{W}^{\text{rep}}\sim f(\boldsymbol{W}^{\text{rep}}\mid\boldsymbol{Y},\boldsymbol{X};\widehat{\psi},\widehat{\theta},\zeta^{W},\zeta^{Y}).
	\end{equation*}
	The following lemma shows asymptotically valid Type I error control under Hypothesis \ref{ci_unobs_hyp} with the true sensitivity parameters.
	\begin{lemma}
		\label{lem:FRT_asmp_unobs}
		Suppose that the joint distribution of observed data admits a density function $f(\boldsymbol{Y},\boldsymbol{W},\boldsymbol{X};\psi,\theta,\zeta^{W},\zeta^{Y})=f(\boldsymbol{W}\mid\boldsymbol{Y},\boldsymbol{X};\psi,\theta,\zeta^{W},\zeta^{Y})f(\boldsymbol{Y},\boldsymbol{X})$, where $f(\boldsymbol{W}\mid\boldsymbol{Y},\boldsymbol{X};\psi,\theta,\zeta^{W},\zeta^{Y})$ defined as \eqref{post_pred_W} is continuous at $\left(\psi_0,\theta_0\right)\in\Psi\times\Theta$. If Hypothesis \ref{ci_unobs_hyp} holds with $\left(\zeta^{W},\zeta^{Y}\right)=\left(\zeta^{W}_0,\zeta^{Y}_0\right)$, and if $\left(\widehat{\psi},\widehat{\theta}\right)\xrightarrow{p} \left(\psi_0,\theta_0\right)$, then
		\begin{equation*}
			\limsup_{n\rightarrow\infty} \Pr(\widehat{p}(\zeta^{W}_0,\zeta^{Y}_0) \leq \alpha) \leq \alpha\quad\text{ for }\alpha \in (0, 1),
		\end{equation*}
		where the probability is taken over the joint distribution $f(\boldsymbol{Y},\boldsymbol{W},\boldsymbol{X};\psi_0,\theta_0,\zeta^{W}_0,\zeta^{Y}_0)$.
	\end{lemma}
	For Bayesian inference, let $f(\psi\mid\boldsymbol{W},\boldsymbol{X};\zeta^{W})$ and $f(\theta\mid\boldsymbol{Y},\boldsymbol{X};\zeta^{Y})$ be posterior distributions derived from \eqref{marginal_W} and \eqref{marginal_Y}, respectively. We calculate Fisher's (or posterior predictive) $p$-value $\widehat{p}_{\text{pos}}(\zeta^{W},\zeta^{Y})$ using the reference distribution $$\boldsymbol{W}^{\text{rep}}\sim \int f(\boldsymbol{W}^{\text{rep}}\mid\boldsymbol{Y},\boldsymbol{X};\psi,\theta,\zeta)f(\psi\mid\boldsymbol{W},\boldsymbol{X};\zeta^{W})f(\theta\mid\boldsymbol{Y},\boldsymbol{X};\zeta^{Y})d\theta d\psi.$$ The frequentist Type-I error control for $\widehat{p}_{\text{pos}}(\zeta^{W}_0,\zeta^{Y}_0)$ can be established similarly to $\widehat{p}(\zeta^{W}_0,\zeta^{Y}_0)$ in combination with the Bernstein-von Mises theorem.
	
	In practice, the true sensitivity parameters are unknown. Our sensitivity analysis involves calculating Fisher's $p$-values over a range of plausible sensitivity parameters to assess how strong unobserved confounding would need to overturn the conclusion drawn assuming no confounding. However, it is computationally difficult as each $p$-value involves thousand times of repeated assignment.
	
	We propose a computationally efficient algorithm to estimate a general Fisher's $p$-value function $p(\zeta)$, where the reference distribution depends on $\zeta$ as
	\begin{equation}
		\label{general_W}
		\boldsymbol{W}^{\text{rep}}\sim f(\boldsymbol{W}^{\text{rep}}\mid\boldsymbol{Y},\boldsymbol{X};\zeta).
	\end{equation}
	Let $\mathcal{D}$ be a bounded open set in the $d$-dimensional Euclidean space $\mathcal{R}^{d}$ and $\{\zeta_m\}_{m=1}^M$ be an equally spaced grid of $M$ points spanning $\mathcal{D}$. Let $I(A)$ denote the indicator function for a general event $A$, which takes $1$ on $A$ and $0$ otherwise. For each $m$-th grid point $\zeta_m$, we find a binary indicator $y_m=I(T_{\text{pos}}(\boldsymbol{Y},\boldsymbol{W}^{\text{rep}}_m,\boldsymbol{X})\geq T_{\text{pos}}(\boldsymbol{Y},\boldsymbol{W},\boldsymbol{X}))$, where $\boldsymbol{W}^{\text{rep}}_m$ is drawn from \eqref{general_W} with $\zeta=\zeta_m$. We then estimate $p(\zeta)$ using the Nadaraya-Watson estimator \citep{nadaraya1964estimating,watson1964smooth} applied to $\{(\zeta_m, y_m)\}_{m=1}^M$ as
	\begin{equation}
		\label{NW}
		\widetilde{p}(\zeta) = \frac{\sum_{m=1}^{M} y_m K_h(\zeta - \zeta_m)}{\sum_{m=1}^{M} K_h(\zeta - \zeta_m)},
	\end{equation}
	where $K_h(u) = h^{-d}K(u/h)$ is the scaled kernel function with bandwidth $h$. The following proposition establishes the consistency of this estimator.
	
	\begin{proposition}
		\label{proposition:estimation}
		Suppose that $p(\zeta)$ is continuous at $\zeta_0\in\mathcal{D}\subset\mathcal{R}^{d}$. Let $K$ be a Riemann integrable kernel function with compact support, satisfying $\int K(u)du=1$ and $\int K^2(u)du<\infty$. Also, let the bandwidth $h$ satisfy $h \to 0$ and $Mh^d \to \infty$ as $M \to \infty$.
		Then $\widetilde{p}(\zeta_0)$ converges in probability to $p(\zeta_0)$ as $h \to 0,M \to \infty$.
	\end{proposition}
	
	\section{Application to a Realistic Example}
	\label{sec:application}
	In this section, we apply the FRT framework and sensitivity analysis to re-examine a realistic dataset from \cite{imbens2001estimating}, who analyzed the effects of unearned income on labor earnings, savings, and consumption using a survey of lottery players. In this study, a linear outcome model is specified under SUTVA for unit $i$ in post-lottery year $j$ as
	\begin{equation}
		\label{structural}
		Y_{ij}=a_j+W_ib_j+X_id_j+\varepsilon_{ij},\quad1\leq i\leq n,0\leq j\leq6,
	\end{equation}
	where $W_i$ is the yearly prize, $X_i$ are additional covariates, $a_j,b_j,d_j$ are coefficients, and $\varepsilon_{ij}$ is a random error. The coefficient $b_j$ (a product of two parameters in their original model) represents the effect of interest in the $j$-th post-lottery year. \cite{imbens2001estimating} use standard ordinary least square (OLS) estimators, where a critical assumption for OLS is the random assignment of prize magnitudes. However, they note that the chance of winning a major prize appears to depend on covariates, raising concerns about unobserved confounding. Another potential source of unobserved confounding comes from non-respondents, who are discarded in this study.
	
	Let $\boldsymbol{Y}_j$ denote the outcome vector for the $j$-th post-lottery year. We first build test statistics for testing $$H_{0j}:\boldsymbol{Y}_j\ci\boldsymbol{W}\mid\boldsymbol{X},\quad 0\leq j\leq 6.$$ Suppose that $\varepsilon_{ij}\sim N(0,\sigma_{j}^2)$ in the outcome model \eqref{structural}, $H_{0j}$ implies $b_j=0$. We specify a non-informative prior as $\pi(a_j,b_j,d_j,\sigma_j)\propto\sigma_j^{-2}$, and derive the test statistic as
	\begin{equation*}
		\begin{aligned}
			T_{\text{pos},j}&=-\operatorname{log}(f(b_j=0\mid\boldsymbol{Y}_j,\boldsymbol{W},\boldsymbol{X}))\\
			&=\operatorname{log}\left(\widehat{\operatorname{se}}(b_j)\right)+\frac{n-k-1}{2}\operatorname{log}\left(1+\widehat{b}_j^2/\left((n-k-1)\widehat{\operatorname{se}}(b_j)^2\right)\right)+\text{constant},
		\end{aligned}
	\end{equation*}
	where $\widehat{b}_j,\widehat{\operatorname{se}}(b_j)$ are OLS estimators for $b_j$ and its standard deviation, respectively. The constant term does not depend on $\boldsymbol{W}$ and is dropped in subsequent calculations.
	
	Next, we build an assignment model to generate the reference distribution for $T_{\text{pos},j}$ under $H_{0j}$. Let $K_i$ be a binary indicator for winning the yearly prize $(K_i=1, W_i>0)$ or not $(K_i=W_i=0)$. Our assignment model extends that of \cite{keisuke2004continuous} as
	\begin{equation}
		\label{structural1}
		\begin{aligned}
			K_i&=I(a_{k}+X_id_{k}+\xi_i>0),\\
			\operatorname{log}(W_i)&=\begin{cases}
				a_{w}+X_id_{w}+\nu_i, \quad\text{if }K_i=1,\\
				-\infty, \quad\text{if }K_i=0,
			\end{cases}
		\end{aligned}
	\end{equation}
	where $\xi_i\sim N(0,1)$, $\nu_i\sim N(0,\sigma_w^2)$. Let $\widehat{\psi}=\left(\widehat{a}_{k},\widehat{d}_{k},\widehat{a}_{w},\widehat{d}_{w},\widehat{\sigma}_w^2\right)$ denote the estimated parameters obtained by fitting a probit model for $K_i$ and an OLS model for $\operatorname{log}(W_i)$ conditional on $K_i=1$. To generate a replicate assignment, we first simulate $K_i^{\text{rep}} = I(\widehat{a}_{k}+X_i\widehat{d}_{k}+\xi_i^{\text{rep}}>0)$ with $\xi_i^{\text{rep}}\sim N(0,1)$. Then, if $K_i^{\text{rep}}=1$, we simulate $\operatorname{log}(W_i^{\text{rep}}) = \widehat{a}_{w}+X_i\widehat{d}_{w}+\nu_i^{\text{rep}}$ with $\nu_i^{\text{rep}}\sim N(0,\widehat{\sigma}_w^2)$; otherwise, $W_i^{\text{rep}}=0$. This process yields $\boldsymbol{W}^{\text{rep}} \sim f(\boldsymbol{W}^{\text{rep}}\mid\boldsymbol{X}; \widehat{\psi})$. Using specification IV of \cite{imbens2001estimating}, which involves the full sets of units and covariates, Table \ref{tab:fep} shows Fisher's $p$-values for testing $H_{0j}$ ($0\leq j\leq 6$). The results suggest that all hypotheses are rejected at the significance level $\alpha=0.05$ under the assumption of no unobserved confounding.
	
	\begin{table}[htbp]
		\centering
		\caption{Fisher's $p$-Values for the Effect of Lottery Winnings on Earnings by Post-Lottery Year}
		\label{tab:fep}
		\begin{tabular}{lccccccc}
			\toprule
			& Year 0 & Year 1 & Year 2 & Year 3 & Year 4 & Year 5 & Year 6 \\
			\midrule
			Fisher's $p$-value & $0.0001$ & $0.0000$ & $0.0000$ & $0.0000$ & $0.0000$ & $0.0000$ & $0.0000$       \\
			\bottomrule
		\end{tabular}
	\end{table}

	We then conduct a sensitivity analysis to assess the robustness of the results in Table \ref{tab:fep} to unobserved confounding. To explain away the causal effect, we introduce an unobserved confounder $\boldsymbol{U}$ into both the outcome and assignment models as
	\begin{equation}
		\label{structural2}
		\begin{aligned}
			Y_{ij}&=a_j+X_id_j+\sigma_{j}\left(\zeta^Y_jU_i+\sqrt{1-(\zeta^Y_j)^2}\varepsilon'_{ij}\right),\\
			K_i&=I\left(a_{k}+X_id_{k}+\zeta^{W}U_i+\sqrt{1-(\zeta^W)^2}\xi'_i>0\right),\\
			\operatorname{log}(W_i)&=\begin{cases}
				a_{w}+X_id_{w}+\sigma_w\left(\zeta^{W}U_i+\sqrt{1-(\zeta^W)^2}\nu'_i\right), \quad\text{if }K_i=1,\\
				-\infty, \quad\text{if }K_i=0,
			\end{cases}
		\end{aligned}
	\end{equation}
	where $U_i,\varepsilon'_{ij},\xi'_i,\nu'_i$ are independent standard normal random variables. The sensitivity parameters $\zeta^Y_j,\zeta^{W}\in[-1,1]$ describe the partial correlations of $U_i$ with the residuals of the outcome and the assignment models, respectively. By integrating out $\boldsymbol{U}$, the conditional distribution for $\boldsymbol{W}$ is given by
	\begin{equation}
		\label{structural3}
		\begin{aligned}
			K_i&=I\left(a_{k}+X_id_{k}+\zeta_j\sigma_j^{-1}\left(Y_{ij}-a_j-X_id_j\right)+\sqrt{1-\zeta^2}\xi'_i>0\right),\\
			\operatorname{log}(W_i)&=\begin{cases}
				a_{w}+X_id_{w}+\sigma_w\left(\zeta_j\sigma_j^{-1}\left(Y_{ij}-a_j-X_id_j\right)+\sqrt{1-\zeta^2}\nu'_i\right), \quad\text{if }K_i=1,\\
				-\infty, \quad\text{if }K_i=0,
			\end{cases}
		\end{aligned}
	\end{equation}
	where $\zeta_j=\zeta^Y_j\zeta^{W}$ describes the partial correlation between $\boldsymbol{W}$ and $\boldsymbol{Y}_j$ conditional on $\boldsymbol{X}$.
	
	By integrating out $\boldsymbol{U}$, the marginal likelihood for $\theta_j=(a_j,d_j,\sigma_j)$ derived from \eqref{structural2} is identical to that derived from \eqref{structural} with $b_j=0$, which allows us to estimate the remaining parameters with OLS as $\widehat{\theta}_j=(\widehat{a}_j,\widehat{d}_j,\widehat{\sigma}_j^2)$. Similarly, the likelihood for $\psi$ derived from \eqref{structural2} is identical to that derived from \eqref{structural1}, which allows us to reuse the estimator $\widehat{\psi}=\left(\widehat{a}_{k},\widehat{d}_{k},\widehat{a}_{w},\widehat{d}_{w},\widehat{\sigma}_w^2\right)$.  To generate a replicate assignment, we first compute the standardized residual as $r_{ij} = (Y_{ij} - (\widehat{a}_j+X_i\widehat{d}_j))/\widehat{\sigma}_j$. Then we simulate $K_i^{\text{rep}} = I(\widehat{a}_{k}+X_i\widehat{d}_{k}+\zeta_j r_{ij}+\sqrt{1-\zeta^2}\xi'_i>0)$, where $\xi'_i \sim N(0,1)$. If $K_i^{\text{rep}}=1$, we simulate $\operatorname{log}(W_i^{\text{rep}}) = \widehat{a}_{w}+X_i\widehat{d}_{w}+\widehat{\sigma}_w\left(\zeta_j r_{ij}+\sqrt{1-\zeta^2}\nu'_i\right)$, where $\nu'_i \sim N(0,1)$; otherwise, $W_i^{\text{rep}}=0$. This process yields $\boldsymbol{W}^{\text{rep}} \sim f(\boldsymbol{W}^{\text{rep}}\mid\boldsymbol{Y}_j, \boldsymbol{X}; \widehat{\theta}_j, \widehat{\psi}, \zeta_j)$.
	
	We consider the Fisher's $p$-value function $\widehat{p}_j(\zeta)$ for sensitivity analysis, where the reference distribution is given above with $\zeta_j=\zeta$. Note that each reference distribution is continuous in its parameters unless degenerated. By \cite{scheffe1947useful}'s theorem, we have the $\widehat{p}_{j}(\zeta)$ for $\zeta\in(-1,1)$. To practically estimate this function, we use the Nadaraya-Watson estimator $\widetilde{p}_j(\zeta)$ defined in \eqref{NW} fitted over an equally spaced grid of points spanning $(-1,1)$. Under the conditions of Proposition~\ref{proposition:estimation}, the Nadaraya-Watson estimator converges in probability to $\widehat{p}_{j}(\zeta)$ on $(-1,1)$.
	
\begin{figure}[htbp]
		\centering
		\caption{Fisher's $p$-Value Functions for the Effect of Lottery Winnings on Earnings by Post-Lottery Year with Unobserved Confounding}
		\label{fig:sensitivity_curves}
		\includegraphics[width=.75\textwidth]{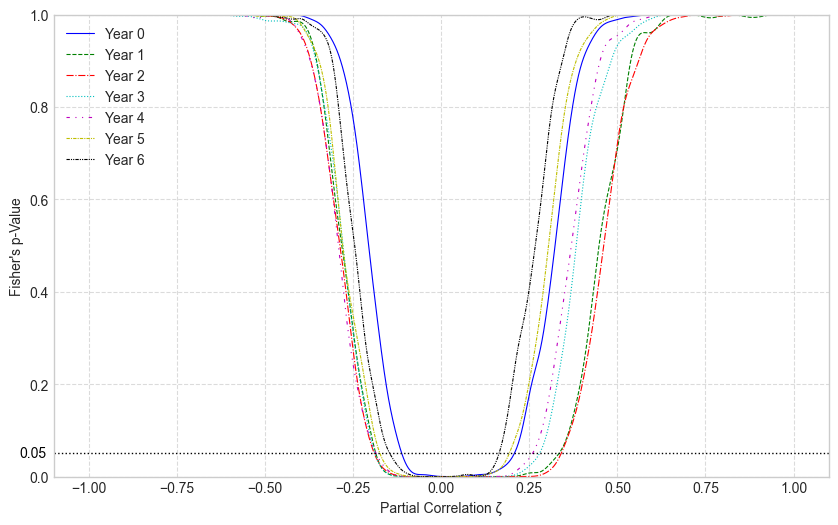}
	\end{figure}

	Figure \ref{fig:sensitivity_curves} shows the estimated $p$-value functions for all post-lottery years $0\leq j\leq6$. Note that all estimated $p$-values at $\zeta=0$ (implying no unconfounding) are close to zero, which is consistent with results in Table \ref{tab:fep}. As $|\zeta|$ increases (implying stronger confounding), the estimated $p$-values become larger.
	For each post-lottery year $j$, we identify the minimum absolute partial correlation $|\zeta_j^*|$ such that $\widetilde{p}_j(\zeta_j^*)\geq0.05$.
	
\begin{figure}[htbp]
		\centering
		\caption{Minimal Strength of Unobserved Confounding to Overturn Conclusions in Table \ref{tab:fep} Across Post-Lottery Years}
		\label{fig:U_curve}
		\includegraphics[width=.75\textwidth]{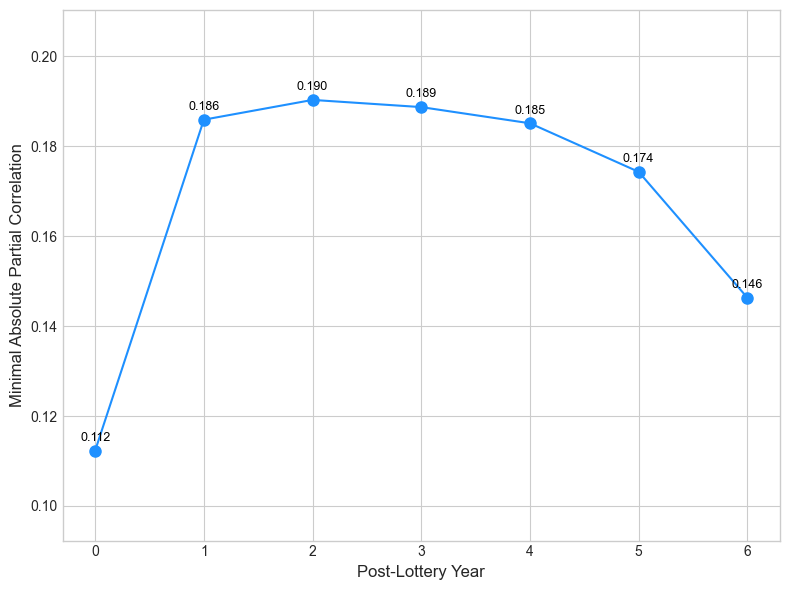} 
	\end{figure}

	Figure \ref{fig:U_curve} shows these minimum absolute partial correlations against post-lottery years. The plot exhibits an inverted U-shape, suggesting dynamic robustness of the causal effect over time. The initial increase in robustness suggests the causal effect of lottery winnings on earnings might take some time to manifest. The subsequent decline in robustness indicates that the causal effect begins to diminish, making it easier for unobserved confounding to explain away the effect.
	
	In order to assess the robustness of these findings to the specific functional forms of the assignment model, we also explore the use of random forests, where trees are honestly built \citep{breiman2001random,athey2019generalized}. The detailed analysis is provided in Supplementary Material.
	
	\section{Discussion and Conclusion}
	\label{sec:conc}
	We have established Fisher's randomization test (FRT) as a design-based inference tool for testing conditional independence between outcomes and treatments given covariates. We provide causal identification of such a hypothesis under the generalized unconfoundedness assumption, which circumvents the need for the no-interference component of SUTVA or the positive overlap assumption. A unique advantage of FRT lies in the distinct roles of assignment and outcome models: a known or consistently estimated assignment mechanism guarantees Type I error control, while a family of outcome models, without being correctly specified, defines Bayesian power and guides the choice of the optimal test statistic.
	
	Our proposed methods are designed to be integrated into a pipeline for observational causal inference. Compared with existing approaches, our pipeline requires a minimal set of presumptions and includes robustness check for possible violation of these presumptions. A direction for future research is to extend this framework to estimate the magnitude of causal effects, and to conduct the sensitivity analysis to assess the robustness of these quantitative estimates to unobserved confounding.

	\section{Disclosure statement}\label{disclosure-statement}
	The authors have no conflicts of interest to declare.
	
	\section{Data Availability Statement}\label{data-availability-statement}
	The data used in this study are from the work of \citet{imbens2001estimating} and publicly available at https://github.com/xuyiqing/lalonde.
	
	\phantomsection\label{supplementary-material}
	\bigskip
	\begin{center}
		{\large\bf SUPPLEMENTARY MATERIAL}
	\end{center}
	
	\begin{description}
		\item[] 
		The PDF file contains all proofs of Lemmas, Theorems, and Propositions from the main paper, as well as an alternative modeling strategy that extends Section \ref{sec:application}.
		\item[]
		The Python code runs the pipelines as described in Section \ref{sec:application} and those extended in Supplementary Material
	\end{description}

	\bibliography{name}
	
	\newpage
	
	\setcounter{section}{0} 
	\setcounter{equation}{0} 
	\setcounter{figure}{0} 
	\setcounter{table}{0} 
	\setcounter{theorem}{0} 
	\setcounter{lemma}{0}   
	\setcounter{proposition}{0} 
	\setcounter{definition}{0} 
	\setcounter{page}{1}
	
	\renewcommand{\thesection}{S\arabic{section}}
	\renewcommand{\theequation}{S\arabic{equation}}
	\renewcommand{\thefigure}{S\arabic{figure}}
	\renewcommand{\thetable}{S\arabic{table}}
	\newtheorem{theorem_supp}{Theorem}[section]
	\newtheorem{lemma_supp}[theorem_supp]{Lemma}
	\newtheorem{proposition_supp}[theorem_supp]{Proposition}
	\newtheorem{definition_supp}[theorem_supp]{Definition}
	\renewcommand{\thetheorem}{S\arabic{theorem}}
	\renewcommand{\thelemma}{S\arabic{lemma}}
	\renewcommand{\theproposition}{S\arabic{proposition}}
	\renewcommand{\thedefinition}{S\arabic{definition}}
	\renewcommand{\thehypothesis}{S\arabic{hypothesis}}
	\begin{center}
		{\Huge\bf Supplementary Material}
	\end{center}

	\section{Proofs of Lemmas, Theorems and Propositions}
	\label{sec:proofs_supplementary}
	
	\noindent\textbf{Proof of Lemma \ref{lem:FRT}.}
	Under Hypothesis \ref{ci_hyp}, we have $f(\boldsymbol{Y}\mid\boldsymbol{W},\boldsymbol{X})=f(\boldsymbol{Y}\mid\boldsymbol{X})$. The probability of rejecting $p\leq\alpha$ under Hypothesis \ref{ci_hyp} can be calculated according to
	\begin{align*}
		&\int I(p\leq\alpha)f(\boldsymbol{Y}\mid\boldsymbol{W},\boldsymbol{X})f(\boldsymbol{W}\mid\boldsymbol{X})f(\boldsymbol{X})d\boldsymbol{Y}d\boldsymbol{W}d\boldsymbol{X}\\
		=&\int\int I(p\leq\alpha)f(\boldsymbol{W}\mid\boldsymbol{X})d\boldsymbol{W} f(\boldsymbol{Y},\boldsymbol{X})d\boldsymbol{Y}d\boldsymbol{X}\\
		\leq&\int\alpha f(\boldsymbol{Y},\boldsymbol{X})d\boldsymbol{Y}d\boldsymbol{X}=\alpha,
	\end{align*}
	where the inequality is because $p$ is dominated by a uniform distribution under repeated randomization from $f(\boldsymbol{W}\mid\boldsymbol{X})$ for a fixed $\boldsymbol{Y}$.\hfill$\square$
	
	\noindent\textbf{Proof of Theorem \ref{ci_explanation}.}
	We first show that statement 1 implies statement 2. If $f(\boldsymbol{W}=\boldsymbol{w}\mid\boldsymbol{X})\cdot f(\boldsymbol{W}=\boldsymbol{w}^{\prime}\mid\boldsymbol{X})>0$, statement 1 implies $f(\boldsymbol{Y}\mid\boldsymbol{W}=\boldsymbol{w},\boldsymbol{X})=f(\boldsymbol{Y}\mid\boldsymbol{W}=\boldsymbol{w}^{\prime},\boldsymbol{X})$. By definition \ref{def:potential}, we have $f(\boldsymbol{Y}\mid\boldsymbol{W}=\boldsymbol{w},\boldsymbol{X})=f(\boldsymbol{Y}(\boldsymbol{w})\mid\boldsymbol{W}=\boldsymbol{w},\boldsymbol{X})$ (the same for $\boldsymbol{w}^{\prime}$). Under equation \eqref{unconfoundedness}, we have $f(\boldsymbol{Y}(\boldsymbol{w})\mid\boldsymbol{W}=\boldsymbol{w},\boldsymbol{X})=f(\boldsymbol{Y}(\boldsymbol{w})\mid\boldsymbol{X})$ (the same for $\boldsymbol{w}^{\prime}$). The desired conclusion follows from combining these equations.
	
	We then show that statement 2 implies statement 1. For fixed $\boldsymbol{w}\in\mathcal{W}^{n}$, Equation \eqref{unconfoundedness} implies $f(\boldsymbol{Y}(\boldsymbol{w}),\boldsymbol{W}\mid\boldsymbol{X})=f(\boldsymbol{Y}(\boldsymbol{w})\mid\boldsymbol{X})f(\boldsymbol{W}\mid\boldsymbol{X})$. If $f(\boldsymbol{W}=\boldsymbol{w}\mid\boldsymbol{X})=0$, we have $f(\boldsymbol{Y}(\boldsymbol{w}),\boldsymbol{W}=\boldsymbol{w}\mid\boldsymbol{X})=0$. If $f(\boldsymbol{W}=\boldsymbol{w}\mid\boldsymbol{X})>0$, we have
	\begin{equation*}
		\begin{aligned}
			f(\boldsymbol{Y}\mid\boldsymbol{X})&=\int f(\boldsymbol{Y},\boldsymbol{W}=\boldsymbol{w}'\mid\boldsymbol{X})d\boldsymbol{w}'\\
			&=\int f(\boldsymbol{Y}(\boldsymbol{w})\mid\boldsymbol{X})f(\boldsymbol{W}=\boldsymbol{w}'\mid\boldsymbol{X})d\boldsymbol{w}'\\
			&=f(\boldsymbol{Y}(\boldsymbol{w})\mid\boldsymbol{X})\int f(\boldsymbol{W}=\boldsymbol{w}'\mid\boldsymbol{X})d\boldsymbol{w}'\\
			&=f(\boldsymbol{Y}(\boldsymbol{w})\mid\boldsymbol{X}),
		\end{aligned}
	\end{equation*}
	where the second equality is because of statement 2.
	We conclude that, for either $f(\boldsymbol{W}\mid\boldsymbol{X})=0$ or $f(\boldsymbol{W}\mid\boldsymbol{X})>0$, $f(\boldsymbol{Y},\boldsymbol{W}\mid\boldsymbol{X})=f(\boldsymbol{Y}(\boldsymbol{W})\mid\boldsymbol{X})f(\boldsymbol{W}\mid\boldsymbol{X})=f(\boldsymbol{Y}\mid\boldsymbol{X})f(\boldsymbol{W}\mid\boldsymbol{X})$, or equivalently, $\boldsymbol{Y}\ci\boldsymbol{W}\mid\boldsymbol{X}$.\hfill$\square$
	
	\noindent\textbf{Proof of Theorem \ref{theorem:bayes_factor}.}
	Given $\boldsymbol{Y},\boldsymbol{X}$, let $f(\boldsymbol{Y}\mid\boldsymbol{X};\alpha)$ denote the $(1-\alpha)$-th quantile of $f(\boldsymbol{Y}\mid\boldsymbol{W},\boldsymbol{X},\theta\in\Theta_1)$ under randomization in $\boldsymbol{W}$ drawn from $f(\boldsymbol{W}\mid\boldsymbol{X})$. The measurability of $f(\boldsymbol{Y}\mid\boldsymbol{X};\alpha)$ is a result of \cite{de2023conditional}. Here, we show that $f(\boldsymbol{Y}\mid\boldsymbol{X};\alpha)$ is also integrable. By definition, the probability of $I(f(\boldsymbol{Y}\mid\boldsymbol{W},\boldsymbol{X},\theta\in\Theta_1)\geq f(\boldsymbol{Y}\mid\boldsymbol{X};\alpha))$ is no smaller than $\alpha$ under randomization in $\boldsymbol{W}$ drawn from $f(\boldsymbol{W}\mid\boldsymbol{X})$. So we can write
	\begin{equation*}
		\alpha f(\boldsymbol{Y}\mid\boldsymbol{X};\alpha)\leq\int f(\boldsymbol{Y}\mid\boldsymbol{W},\boldsymbol{X},\theta\in\Theta_1)f(\boldsymbol{W}\mid\boldsymbol{X})d\boldsymbol{W}.
	\end{equation*}
	Because the right-hand side is a conditional density function, we know that $\int f(\boldsymbol{Y}\mid\boldsymbol{X};\alpha)d\boldsymbol{Y}<\infty$.
	
	Let $p_\text{BF}$ denote Fisher's $p$-value using $T_{\text{BF}}$ as the test statistic. The decision rule following Equation \eqref{rd} (Main Paper, for randomized decision rule) based on $p_\text{BF}$ can be rewritten as
	\begin{equation*}
		\delta_{\text{BF}}(\boldsymbol{Y},\boldsymbol{W},\boldsymbol{X})=\begin{cases}
			1, \text{ if }f(\boldsymbol{Y}\mid\boldsymbol{W},\boldsymbol{X},\theta\in\Theta_1)>f(\boldsymbol{Y}\mid\boldsymbol{X};\alpha^{+}),\\
			q, \text{ if }f(\boldsymbol{Y}\mid\boldsymbol{W},\boldsymbol{X},\theta\in\Theta_1)=f(\boldsymbol{Y}\mid\boldsymbol{X};\alpha^{+}),\\
			0, \text{ if }f(\boldsymbol{Y}\mid\boldsymbol{W},\boldsymbol{X},\theta\in\Theta_1)<f(\boldsymbol{Y}\mid\boldsymbol{X};\alpha^{+}).
		\end{cases}
	\end{equation*}
	For any $\delta\in\Delta_{\alpha}$, the construction of $\Delta_{\alpha}$ implies
	$\int\delta(\boldsymbol{Y},\boldsymbol{W},\boldsymbol{X}) f(\boldsymbol{W}\mid\boldsymbol{X}) d\boldsymbol{W} = \alpha$ for fixed $\boldsymbol{Y},\boldsymbol{X}$.
	Applying the Neyman-Pearson lemma, we find that $\Pr(\delta_{\text{BF}}=1\mid\theta\in\Theta_1)=\int\delta_{\text{BF}}f(\boldsymbol{Y}\mid\boldsymbol{W},\boldsymbol{X},\theta\in\Theta_1)f(\boldsymbol{W},\boldsymbol{X})d\boldsymbol{Y}d\boldsymbol{W}d\boldsymbol{X}$ is optimal within $\Delta_{\alpha}$.
	\hfill$\square$
	
	\noindent\textbf{Proof of Proposition \ref{proposition:sufficiency}.}
	Applying a transformation that does not involve $\boldsymbol{W}$, we have
	\begin{equation*}
		\frac{(T_{\text{BF}}+1)f(\boldsymbol{Y}\mid\boldsymbol{X};\theta\in\Theta_0)}{f(\boldsymbol{Y}\mid\boldsymbol{X};\theta=\theta_{0})}=f^{-1}(\theta=\theta_{0}\mid\boldsymbol{Y},\boldsymbol{W},\boldsymbol{X}).
	\end{equation*}
	Taking logarithm on both sides, we have
	\begin{equation*}
		T_{\text{pos}}=\operatorname{log}(T_{\text{BF}}+1)+\operatorname{constant},
	\end{equation*}
	where the constant term does not depend on $\boldsymbol{W}$. Therefore, $T_{\text{pos}}$ is a strictly monotone transformation of $T_{\text{BF}}$.
	\hfill$\square$
	
	\noindent\textbf{Proof of Lemma \ref{lem:FRT_asmp}.}
	Let $p(\psi)$ denote Fisher's $p$-value with probability taken over $\boldsymbol{W}^{\text{rep}}\sim f(\boldsymbol{W}\mid\boldsymbol{X};\psi)$. By \cite{scheffe1947useful}'s theorem, we know that $p(\psi)$ is continuous in $\Psi$. It follows from the continuous mapping theorem that $\widehat{p}$ converges in probability to $p(\psi)$. For $\alpha\in(0,1)$, picking any $\gamma>0$ such that $\alpha+\gamma<1$, we have
	\begin{equation*}
		\limsup_{n\rightarrow\infty}\Pr(\widehat{p}\leq\alpha)\leq\Pr(p(\psi)\leq\alpha+\gamma)\leq\alpha+\gamma,
	\end{equation*}
	where the last inequality comes from Lemma \ref{lem:FRT}. Letting $\gamma\rightarrow0$ completes the proof. \hfill$\square$
	
	\noindent\textbf{Proof of Lemma \ref{lem:FRT_asmp_unobs}.}
	 Let $p(\theta,\psi)$ denote Fisher's $p$-value with probability taken over $\boldsymbol{W}^{\text{rep}}\sim f(\boldsymbol{W}\mid\boldsymbol{Y},\boldsymbol{X};\theta,\psi,\zeta^{Y}_0,\zeta^{W}_0)$. The probability of rejecting $p(\theta_0,\psi_0)\leq\alpha$ can be calculated as follows:
	\begin{align*}
		&\int I(p(\theta_0,\psi_0)\leq\alpha)f(\boldsymbol{Y}\mid\boldsymbol{X},\boldsymbol{U};\theta_0,\zeta^Y_0)f(\boldsymbol{W}\mid\boldsymbol{X},\boldsymbol{U};\psi_0,\zeta^W_0)f(\boldsymbol{X},\boldsymbol{U})d\boldsymbol{Y}d\boldsymbol{W}d\boldsymbol{X}d\boldsymbol{U}\\
		=&\int\int I(p(\theta_0,\psi_0)\leq\alpha)\int f(\boldsymbol{W}\mid\boldsymbol{X},\boldsymbol{U};\psi_0,\zeta^W_0)f(\boldsymbol{U}\mid\boldsymbol{Y},\boldsymbol{X};\theta_0,\zeta^{Y}_0)d\boldsymbol{U}d\boldsymbol{W} f(\boldsymbol{Y},\boldsymbol{X})d\boldsymbol{Y}d\boldsymbol{X}\\
		=&\int\int I(p(\theta_0,\psi_0)\leq\alpha)f(\boldsymbol{W}\mid\boldsymbol{Y},\boldsymbol{X};\psi_0,\theta_0,\zeta^{W}_0,\zeta^{Y}_0)d\boldsymbol{W} f(\boldsymbol{Y},\boldsymbol{X})d\boldsymbol{Y}d\boldsymbol{X}\\
		\leq&\int\alpha f(\boldsymbol{Y},\boldsymbol{X})d\boldsymbol{Y}d\boldsymbol{X}=\alpha,
	\end{align*}
	where the inequality is because $p(\theta_0,\psi_0)$ is dominated by a uniform distribution under repeated randomization from $f(\boldsymbol{W}\mid\boldsymbol{Y},\boldsymbol{X};\psi_0,\theta_0,\zeta^{W}_0,\zeta^{Y}_0)$. The rest of the proof is similar to Lemma \ref{lem:FRT_asmp} and thus omitted. \hfill$\square$
	
	\noindent\textbf{Proof of Proposition \ref{proposition:estimation}.}
	Consider the following decomposition:
	\begin{equation}
		\label{upper_bound}
		|\widetilde{p}(\zeta) - p(\zeta)| \leq \underbrace{\left|\widetilde{p}(\zeta) - \frac{\sum_{m=1}^{M}\widehat{p}(\zeta_m)K_{h}(\zeta-\zeta_m)}{\sum_{m=1}^{M}K_{h}(\zeta-\zeta_m)}\right|}_{\text{the first term}} + \underbrace{\left|\frac{\sum_{m=1}^{M}\widehat{p}(\zeta_m)K_{h}(\zeta-\zeta_m)}{\sum_{m=1}^{M}K_{h}(\zeta-\zeta_m)} - p(\zeta)\right|}_{\text{the second term}}.
	\end{equation}
	The second term in \eqref{upper_bound} represents the bias of the estimator, which converges to $0$ at $\zeta_0$ as $h\rightarrow0$ by the continuity of $p(\zeta)$.
	
	For the first term in \eqref{upper_bound}, let $e_m = y_m - \widehat{p}(\zeta_m)$. Then $e_m$ ($m=1,\ldots,M$) are mutually independent, zero-mean random variables, with $|e_m|\leq1$.
	By Hoeffding's inequality, for any $\epsilon > 0$,
	\begin{equation}
		\label{convegence_in_p}
		\Pr\left(\left|\frac{\sum_{m=1}^{M}e_mK_{h}(\zeta-\zeta_m)}{\sum_{m=1}^{M}K_{h}(\zeta-\zeta_m)}\right| > \epsilon\right) \leq 2\exp\left(-\frac{2\epsilon^2\left(\sum_{m=1}^{M}K_{h}(\zeta-\zeta_m)\right)^2}{\sum_{m=1}^{M}K_{h}^2(\zeta-\zeta_m)}\right).
	\end{equation}
	Let $S$ be the area of $\mathcal{D}$. As $M\rightarrow\infty$, by the definition of Riemann integral, we have $\sum_{m=1}^{M}K_{h}(\zeta-\zeta_m)=\frac{M}{S}\sum_{m=1}^{M}\frac{S}{Mh^d}K(u_m)\rightarrow\frac{M}{S}\int K(u)du=\frac{M}{S}$, and $h^d\sum_{m=1}^{M}K_{h}^2(\zeta-\zeta_m)=\frac{M}{S}\sum_{m=1}^{M}\frac{S}{Mh^d}K^2(u_m)\rightarrow\frac{M}{S}\int K^2(u)du$. Therefore, the right-hand side of \eqref{convegence_in_p} satisfies
	\begin{equation*}
		\lim_{M\rightarrow\infty}2\exp\left(-\frac{2\epsilon^2\left(\sum_{m=1}^{M}K_{h}(\zeta-\zeta_m)\right)^2}{\sum_{m=1}^{M}K_{h}^2(\zeta-\zeta_m)}\right)=\lim_{M\rightarrow\infty}2\exp\left(-\frac{2\epsilon^2Mh^d}{S\int K^2(u)du}\right)=0
	\end{equation*}
	as $Mh^d\rightarrow\infty$. Consequently, the first term in \eqref{upper_bound} converges in probability to 0 at $\zeta_0$ as $Mh^d\rightarrow\infty$. \hfill$\square$
	\section{Alternative Assignment Modeling}
	Continuing Section \ref{sec:application}, we investigate an alternative specification for the assignment models. Assuming no unobserved confounding, the specification \eqref{structural1} employs a probit model for the probability of winning a prize $K_i$ and a linear regression model for the log-prize amount $\operatorname{log}(W_i)$ conditional on winning. With unobserved confounding explaining away the treatment effect, the specification \eqref{structural2} also introduces linear regression models for post-lottery earnings $Y_{ij}$ ($0\leq j\leq6$) assuming no treatment effect.
	
	We replace all these linear parametric models with random forests (RF) for flexible assignment modeling. First, an RF classifier is trained to model the conditional probability of winning a prize $K_i$ given covariates $\boldsymbol{X}_i$. The estimated probability, $\hat{p}_K(\boldsymbol{X}_i)$, is transformed using the inverse of $\Phi$ to replace the linear component in the original probit model, where $\Phi$ denotes the cumulative function of the standard normal distribution. Second, for units that won a prize in the observed data, an RF regressor is trained to estimate the conditional mean of the log-prize amount $\operatorname{log}(W_i)$ given covariates $\boldsymbol{X}_i$, which replaces the linear conditional mean in the regression model for $\operatorname{log}(W_i)$. Third, for each post-lottery year $j$ ($0\leq j\leq6$), a separate RF regressor is trained to estimate the conditional mean of the post-lottery earning $Y_{ij}$ given covariates $\boldsymbol{X}_i$, which replaces the linear conditional mean in the regression model for $Y_{ij}$. These models are implemented using the EconML Python package \citep{econml}.
	
	The rest of the FRT and sensitivity analysis procedures remain the same as described in Section \ref{sec:application}. As shown below, the results from the RF-based specification do not substantively change our conclusions in Section \ref{sec:application}, suggesting they are robust to misspecification in assignment modeling.
	\begin{table}[htbp]
		\centering
		\caption{Fisher's $p$-Values for the Effect of Lottery Winnings on Earnings by Post-Lottery Year Based on RF Assignment Modeling}
		\label{tab:fep_sup}
		\begin{tabular}{lccccccc}
			\toprule
			& Year 0 & Year 1 & Year 2 & Year 3 & Year 4 & Year 5 & Year 6 \\
			\midrule
			Fisher's $p$-value & $0.0005$ & $0.0000$ & $0.0001$ & $0.0001$ & $0.0000$ & $0.0001$ & $0.0005$       \\
			\bottomrule
		\end{tabular}
	\end{table}
	\begin{figure}[htbp]
		\centering
		\caption{Fisher's $p$-Value Functions Based on RF Assignment Modeling for the Effect of Lottery Winnings on Earnings by Post-Lottery Year with Unobserved Confounding}
		\label{fig:sensitivity_curves_sup}
		\includegraphics[width=.75\textwidth]{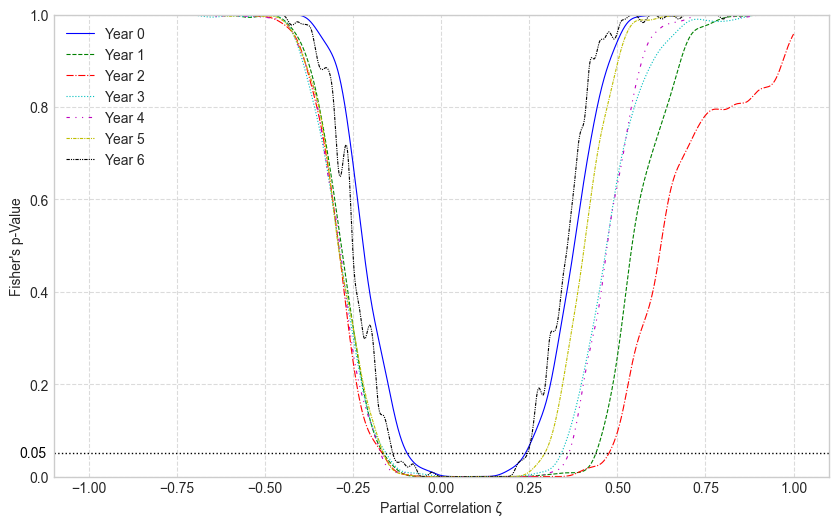}
	\end{figure}
	
	\begin{figure}[htbp]
		\centering
		\caption{Minimal Strength of Unobserved Confounding to Overturn Conclusions in Table \ref{tab:fep_sup} Across Post-Lottery Years}
		\label{fig:U_curve_sup}
		\includegraphics[width=.75\textwidth]{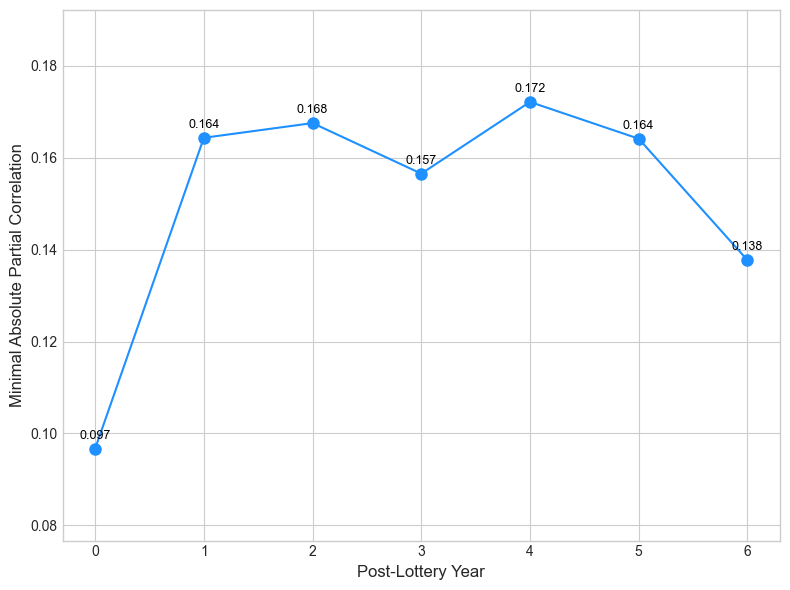} 
	\end{figure}

\end{document}